# CENTRALLY CONDENSED COLLAPSE OF STARLESS CORES

Philip C. Myers
Harvard-Smithsonian Center for Astrophysics
60 Garden Street, Cambridge MA 02138; pmyers@cfa.harvard.edu

ABSTRACT

Models of self-gravitating gas in the early stages of pressure-free collapse are compared for initial states which are equilibrium layers, cylinders, and "Bonnor-Ebert" spheres. For each geometrical case the density profile has an inner region of shallow slope surrounded by an outer region of steep slope, and the profile shape during early collapse remains similar to the profile shape of the initial equilibrium. The two-slope density structure divides the spherical collapse history into a starless infall phase and a protostellar accretion phase. The similarity of density profiles implies that Bonnor-Ebert fits to observed column density maps may not distinguish spherical cores from oblate or prolate cores, and may not distinguish static cores from collapsing cores. The velocity profiles discriminate better than the density profiles between initial geometries and between collapse ages. The infall velocity generally has a subsonic maximum value, which is approximately equal to the initial velocity dispersion times the ratio of collapse age to central free-fall time.

Observations of starless core line profiles constrain collapse models. Collapse from initial states which are strongly condensed and slightly prolate is consistent with "infall asymmetry" observed around starless cores, and is more consistent than collapse from initial states which are weakly condensed, and/or oblate. Spherical models match observed inward speeds 0.05-0.09 km s$^{-1}$ over 0.1-0.2 pc, if the collapse has a typical age 0.3-0.5 free fall times, and if it began from a centrally condensed state which was not in stable equilibrium. In a collapsing core, optically thin line profiles should broaden and develop two-peak structure as seen in L1544, once the typical infall velocity approaches the molecular velocity dispersion, or when the collapse age exceeds $\sim$ 0.4 free fall times, for typical parameters, independent of depletion.

*Subject headings:* stars: formation -- ISM: clouds

1.  INTRODUCTION

Dense cores of interstellar gas (Myers and Benson 1983, Benson & Myers 1989; Cernicharo 1991; Andre′, Ward-Thompson & Barsony 2000) are the formation sites of stars in the nearest star-forming complexes (Beichman et al 1986). Their motions are therefore important for understanding the physics of star formation. Yet the mechanisms of core condensation and star-forming collapse are still unclear. For example the motions in the starless core L1544 have been described as inconsistent with any standard model of star formation (Tafalla et al 1998), but they have also been attributed to contraction due to turbulent dissipation (Myers & Lazarian 1998), to a rotating, infalling envelope (Ohashi et al 1999), and to radial inward motions in a disk-like structure undergoing magnetically diluted collapse and ambipolar diffusion (Ciolek & Basu 2000). In this paper we compare models of collapsing systems, to discriminate among degrees of central condensation and among simple geometrical shapes. We discuss one model, the collapsing Bonnor-Ebert sphere, in more detail, and show that it can match many observed properties of starless cores.

Recent observations have clarified the following properties of the brightest starless cores. Many of these cores detected and mapped at submillimeter wavelengths are also known as "prestellar" or "pre-protostellar" cores because of the expectation that they are close to the point of star formation (Ward-Thompson, Motte, & Andre′ 1999).

1. *Central condensation*. Observations of submillimeter dust emission and near-infrared absorption of background stars indicate that many low-mass cores are strongly centrally condensed, with nearly constant density ~ 3-10 × $10^5$ $cm^{-3}$ within a few $10^3$ AU, surrounded by a steeply declining power-law envelope (Ward-Thompson et al 1994, André, Ward-Thompson & Motte 1996). When they are modelled as arising from a self-gravitating isothermal sphere of constant mass in hydrostatic equilibrium (Bonnor 1956, Ebert 1955), the best fit to the data indicates a sphere which is strongly centrally condensed (Evans et al 2001, Johnstone et al 2001). The cores B68, L694-2, and several cores in NGC 2068 appear too strongly condensed to be in stable equilibrium (Alves, Lada & Lada 2001, Johnstone et al 2001, Harvey et al 2003). However, the spherical models cannot be completely accurate, since the typical map shows elongation with aspect ratio 1-2.

2. *Inward motions*. In nearby regions of low-mass star formation, some 20 starless cores show evidence of "infall asymmetry" in spectral lines of CS, HCO$^+$, DCO$^+$, and N$_2$H$^+$ (Tafalla et al 1998, Lee, Myers & Tafalla 1999, Onishi, Mizuno & Fukui 1999, Gregersen & Evans 2000, Lee, Myers & Tafalla 2001, Caselli et al 2002a, Lee, Myers & Plume 2004, Bourke et al 2004, Crapsi et al 2004). These observations indicate the typical projected extent of the inward motions, in the CS 2-1 line, is 0.2-0.3 pc. This extent exceeds that of the dense core map, as indicated by the FWHM of the integrated intensity of the 1-0 line of N$_2$H$^+$, by a factor of 2-3. The effective infall speed, defined as the line-of-sight speed with which two layers approach their common center in a plane-parallel radiative transfer model (Myers et al 1996), is typically 0.05-0.09 km s$^{-1}$ (Lee et al 2001). In a few well-studied cores, the infall speed toward the position of peak emission appears to increase inward, as in L1544 and L694-2 (Bourke et al 2004), and in L183, L1521F, and L1689B (Lee, Myers & Plume 2004). In L1544 and L694-2 this increase is from a few 0.01 km s$^{-1}$ to $\sim$ 0.1 km s$^{-1}$ (Bourke et al 2004).

3. *Nearly uniform gas temperature*. In L1517 and L1498, high-resolution observations of the (J, K) = (1,1) and (2,2) lines of NH$_3$ indicate relatively uniform kinetic temperature 10 ± 1 K over each core (Tafalla et al 2004), confirming estimates based on lower-resolution observations (Benson & Myers 1989). These results for the gas differ from temperature estimates of the dust, which indicate cooler temperatures in the core interior, according to far infrared color temperature maps (Ward-Thompson, Andre$'$ & Kirk 2002) and according to models in which core dust heating is dominated by external starlight (Evans et al 2001, Zucconi, Walmsley & Galli 2001). When thermal coupling between gas and grains is taken into account, the gas is expected to be cooled by the dust for gas density greater than $\sim$ 10$^5$ cm$^{-3}$, but even at 10$^6$ cm$^{-3}$, the predicted gas temperature does not fall below about 8 K, regardless of the degree of molecular depletion due to freeze-out on grains (Goldsmith 2001).

4. *Subsonic turbulence*. The motions of the "inner core" which appear neither thermal nor inward are subsonic, with magnitude at most 0.05-0.1 km s$^{-1}$. These motions are evident as residual contributions to the line width (Myers 1983, Fuller & Myers 1993, Goodman et al 1998), and as variations in the center velocity of the N$_2$H$^+$ 1-0 line from point to point in a map. In some cores such as L1512 these variations are smooth gradients suggestive of rotation, and in others such as L1521F the velocity patterns are more complex (Caselli et al 2002a,b, Tafalla et al 2002, Crapsi et al 2004, Tafalla et al 2004).

5. *Marginal magnetic fields.* Magnetic fields are expected to influence core formation and evolution (Shu, Adams & Lizano 1987), but their importance for dynamics of well-studied cores is not clear from the available observations. Submillimeter continuum polarization directions are not aligned with the projected axes of L1544, L183, and L43 (Ward-Thompson et al 2000, Crutcher et al 2004), suggesting that magnetic forces do not dominate their structure, at least for simple core shapes. In L1544, the mass-to flux ratio is apparently not strongly subcritical, as would be required for magnetic forces to dominate the recent evolution of the core. Instead the mass-to-flux ratio appears nearly critical, according to an ambipolar diffusion model which matches observed densities and velocities (Ciolek & Basu 2000), or slightly supercritical, based on analysis of the polarization directions (Crutcher et al 2004), or on analysis of OH Zeeman observations (Crutcher & Troland 2000). Observations of more cores are needed for a more definite conclusion.

6. *Selective depletion.* High-resolution maps of line emission from isotopomers of $N_2H^+$ and $NH_3$ closely match dust emission maps, while maps of emission from isotopomers of CO, CS, $H_2CO$, and other species in dense cores have "holes," "valleys" or "plateaus" at the dust emission peak (Caselli et al 1999, Tafalla et al 2002), in accord with models of molecular freeze-out onto grain surfaces for gas densities greater than $3\text{-}10 \times 10^5$ cm$^{-3}$ (Bergin & Langer 1997). Similarly, deuterated species show enhanced abundance correlated with the depletion of CO (Crapsi et al 2004).

These properties suggest that models of core motions should focus on systems with strong central condensation, modest projected elongation, and nearly isothermal gas pressure, in which turbulent motions and magnetic effects are secondary. This picture is similar to that of earlier equilibrium models of low-mass dense cores (Shu 1977, Myers & Fuller 1992), but with better definition of the density and depletion structure.

The three-dimensional geometry of dense cores is evidently not spherical, but it is not settled whether an oblate, prolate, triaxial, or other description is most appropriate to match the typical 2:1 aspect ratio of a core map. If the choice is confined to oblate or prolate shapes, then statistical arguments favor prolate shapes, especially in cases where a core is the densest part of a filamentary cloud (Myers et al 1991, Ryden 1996). More complex shapes such as toroids (Li & Shu 1996) or nearly oblate triaxial ellipsoids (Jones, Basu & Dubinski 2001) may also be consistent with observations. In this paper we consider collapsing motions arising from centrally condensed isothermal systems which are extremely prolate, i.e. a cylinder of constant mass per unit length; extremely oblate, i.e. a layer of constant mass per unit area, and spherical, i.e. the sphere of constant mass (the "Bonnor-Ebert" sphere).

The nonsingular centrally condensed initial state considered here is distinctly different from initial states which are uniform, or infinitely centrally condensed, and the resulting collapse profiles of velocity and density are also distinctly different from those which arise from such initial states. The calculation most similar to those given here is for spherical free-fall starting from a centrally condensed "Plummer-like" density law (Whitworth and Ward-Thompson 2001, hereafter WW), and our results agree well with theirs. As in the WW model, the density profile of the isothermal equilibrium state is specified by two parameters, e.g. the central density and the temperature. A third parameter is needed in each case to specify the outer boundary. Unlike the WW model, the isothermal equilibrium state is a physically well-defined system in force balance.

We consider pressure-free collapse rather than the more realistic isothermal collapse, because it is much easier to compare a variety of pressure-free collapses, which have analytic solutions. The isothermal collapse of a centrally condensed sphere with finite central density has been treated in numerical simulations (Foster & Chevalier 1993, Ogino, Tomisaka, & Nakamura 1999, Motoyama & Yoshida 2003, Aikawa et al 2004), and isothermal collapse of a singular isothermal sphere has been analyzed by similarity solutions (Shu 1977, Hunter 1977, Whitworth & Summers 1985). Our pressure-free collapse results share many features with the available results for isothermal collapses, as discussed in Section 5.3.

Initial isothermal equilibrium solutions are well known for layers (Spitzer 1942), cylinders (Stodolkiewicz 1963, Ostriker 1964), and spheres (Chandrasekhar 1939; Spitzer 1965; Bonnor 1956; Ebert 1955). The free-fall collapse solutions are given for spheres by Hunter (1962), and for layers and cylinders by Penston (1969). Although these solutions are relatively simple to evaluate, to our knowledge no detailed comparison has been presented of collapse solutions from centrally condensed states in different geometries.

In the following models, the collapse is assumed to arise from "sudden cooling" in which the time scale of thermal cooling is short compared to the dynamical time. The temperature drops from its initial equilbrium value $T_i$ to a value T low enough so that the thermal pressure gradient opposing self-gravity is negligible. This choice for inducing collapse is motivated by observations which indicate that dense cores have much less turbulence than their surrounding envelopes (e.g. Goodman et al 1998), and by simulations which show that turbulent motions in molecular clouds dissipate rapidly (e.g., MacLow & Klessen 2004).

We present the collapse equations in Section 2, and results in Section 3. In Section 4 we apply these results to a comparison with observations of infall asymmetry in starless cores. We conclude with a discussion and summary in Section 5.

## 2. COLLAPSE EQUATIONS

### 2.1. Spherical collapse

For a sphere of fixed mass, surrounded by an infinite medium of constant pressure, initially in isothermal equilibrium with velocity dispersion $\sigma$ (a "Bonnor-Ebert" sphere, Bonnor 1956, Ebert 1955) we combine the equilibrium expressions in Spitzer (1965) and the collapse expressions in Hunter (1962), whence at time t we have for velocity v, radius r, and density n,

$$v = -\sigma \tan\beta \, (2\psi' \, \xi)^{1/2} \tag{1}$$

$$r = (2/\pi)(2/3)^{1/2} \, \sigma \, t_f \xi \, \cos^2\beta \tag{2}$$

$$n = n_0 \exp(-\psi) \sec^6\beta \tag{3}$$

where the central free fall time of the sphere is

$$t_f = [3\pi/(32Gmn_0)]^{1/2} \tag{4}$$

with $n_0$ equal to the initial central density and where the parameter $\beta$ is related to the initial structure and the elapsed time by

$$\beta + (1/2)\sin 2\beta \equiv \gamma = (\pi/2)(t/t_f)(3\psi'/\xi)^{1/2} \tag{5}$$

Here $\psi$ and $\psi'$ are functions of the dimensionless initial radius $\xi$ given by Chandrasekhar and Wares (1949); $\xi$, $\psi$ and $\psi'$ are defined by

$$\xi \equiv a/a_0 \equiv a(4\pi Gmn_0)^{1/2}/\sigma \tag{6}$$

$$\psi \equiv -\ln(n/n_0) \tag{7}$$

$$\psi' \equiv d\psi/d\xi \tag{8}$$

Here a is the initial radius of the mass shell with current radius r, and $a_0$ is the scale length, comparable to the "flat radius" in the density profile. The parameters $\beta$ and $\gamma$ increase with time and decrease with increasing radius.

It is useful to invert eq. (5) so that $\beta$ can be expressed in terms of $\gamma$. For this purpose we have found the approximation

$$\beta = (1/2)\gamma(1.585-\gamma)^{-0.16} \tag{9}$$

accurate to better than 7% over the range $0 \le \gamma \le \pi/2$. The calculation proceeds by listing $\xi$, $\psi$, and $\psi'$ for $\xi$ ranging from 0 to $\xi_1$, then for given $t/t_{f,s}$ obtaining $\gamma$ from eq. (5), $\beta$ from eq. (9), and then v, r, and n from eqs. (1-3).

2.2. Cylindrical collapse

For the infinite cylinder collapsing radially from initial isothermal equilibrium, we combine the equilibrium expressions in Ostriker (1964; see also Stodolkiewicz 1963) and the collapse expressions in Penston (1969). Then

$$v_c = -8^{1/2}\sigma\,\theta\,[1-(\mathrm{erf}(\theta)/(t/t_{f,c}))^2]^{1/2} \tag{10}$$

$$r_c = (8/\pi)^{1/2}\sigma\, t_{f,c}\exp(-\theta^2)\,[((t/t_{f,c})/\mathrm{erf}(\theta))^2 - 1]^{1/2} \tag{11}$$

$$n_c = n_0\exp(2\theta^2)[\mathrm{erf}(\theta)/(t/t_{f,c})]^4 \tag{12}$$

where the central free fall time is

$$t_{f,c} = [1/(4Gmn_0)]^{1/2} \tag{13}$$

and where we use the subscript "c" to distinguish the velocity, radius, and density of the cylinder from those of the sphere. Here $\theta$ is a parameter which increases from 0 to $\theta_{max}$, where $\mathrm{erf}(\theta_{max}) = t/t_{f,c}$. The calculation proceeds by listing $\theta$ and

erf($\theta$) for $\theta$ ranging from 0 to $\theta_{max}$, and then obtaining $v_c$, $r_c$, and $n_c$ from eqs. (10-12).

2.3. Planar collapse

For the infinite planar layer collapsing from initial isothermal equilibrium, we combine the equilibrium expressions in Spitzer (1942) and the collapse expressions in Penston (1969). These give

$$v_z = -2\sigma (t/t_{f,l}) \tanh(z_i/(\sigma t_{f,l})) \qquad (14)$$

$$z = z_i - \sigma t(t/t_{f,l}) \tanh(z_i/(\sigma t_{f,l})) \qquad (15)$$

$$n_l = n_0 [\cosh^2(z_i/(\sigma t_{f,l})) - (t/t_{f,l})^2]^{-1} \qquad (16)$$

where the central free fall time is

$$t_{f,l} = [1/(2\pi G m n_0)]^{1/2} \qquad (17)$$

Here $z_i$ is the initial height of the slice whose height at time t is z, and $\sigma t_{f,l}$ is the equilibrium scale height of the layer. The calculation is done by listing $z_i$ from 0 to arbitrarily large values, and then evaluating eqs. (14-16) for a chosen value of $t/t_{f,l}$.

3. MODEL RESULTS

In this section we give velocity and density profiles for collapsing systems, based on the equations of Section 2. For sufficiently early times, the velocity profile always has a subsonic maximum value and the density profile always has a central region of shallow slope. The subsonic maximum velocity implies that spectral line shapes should not have "infall wings" and the flat inner density profile implies that a collapsing core has distinct phases of infall and accretion.

3.1. Velocity structure

Figure 1 compares velocity structure at $t/t_f = 1/2$ for a cylinder, a layer, and a sphere. All systems have the same initial equilibrium temperature, 15 K, and central density, $n_0 = 3 \times 10^5$ cm$^{-3}$. The layer and cylinder models extend to infinity, unlike the BE sphere which has a finite boundary. We show all three velocity profiles over the same range of distances as for the sphere, chosen here to have a dimensionless bounding radius $\xi_1 = 10$. Figure 1 indicates that the maximum velocities for the cylinder and layer exceed that of the sphere, but we note that the infinite extent of the cylinder along its axis, and of the layer along its midplane, causes each maximum velocity to exceed the value expected for a figure of finite extent. A rough estimate indicates that the maximum velocity is about the same for all three shapes, if they have the same maximum dimensions.

In this section we give a useful approximation for the maximum velocity. Evaluation of eqs. (1) and (2) indicates that the maximum velocity increases with time, and the position of maximum velocity moves inward with time. The maximum velocity becomes equal to the sound speed at $t/t_f=0.83$, at position $r/a_0=1.6$. For earlier times, the maximum velocity is always subsonic, and can be estimated from eqs. (1) and (5), which indicate that

$$v = - 6^{1/2} (\pi/4) F(\beta)(t/t_f) \sigma \, \psi' \qquad (18)$$

where

$$F(\beta) \equiv 2/[(\beta/\tan \beta) + \cos^2 \beta] \qquad (19)$$

F is a function which approaches unity as $\beta$ approaches 0, and which stays within 10% of unity for $\beta < 0.38$. Thus when $\beta$ is small, both $F(\beta)$ in eq. (18) and $\cos^2 \beta$ in eq. (2) are nearly independent of position, so the velocity is simply proportional to $\psi'$, which has maximum value 0.517 at $\xi = 3.0$ (Chandrasekhar & Wares 1949). In this approximation the maximum velocity is obtained from eq. (18) as

$$v_{max, \, app} = -1.0(t/t_f)\sigma \qquad (20)$$

Eq. (20) indicates that the infall velocity is subsonic with respect to the initial velocity dispersion. The maximum velocity occurs at the position corresponding to $\xi=3.0$, and since we have assumed $\cos^2\beta \approx 1$, eq. (2) yields the approximate radius of maximum velocity as

$$r_{app}(v_{max}) = 1.6 a_0 \qquad (21)$$

This position of the velocity maximum lies just outside the "flat" part of the density profile, as was noted by Foster & Chevalier (1993) in their numerical simulations of isothermal collapse. At $t=t_f/2$, $v_{max, \, app}$ in eq. (20) is 5% low and $r_{app}(v_{max})$ in eq. (21) is 13% high with respect to the exact values from the equations in Section 2.1.

A similar treatment of the velocity expressions for the cylinder and the layer indicates that the maximum velocity is again of the same form as for the sphere, with slightly greater coefficients than in eq. (20). For the layer this result is evident by inspection of eq. (14) if the initial height $z_i$ of a slice lies many scale heights above the midplane: then $\tanh(z_i/\sigma\, t_{f,l})$ approaches unity and $v_{z,max} = 2(t/t_{f,l})\sigma$.

## 3.2. Infall wings

Section 3.1 shows that the infall speed has a subsonic maximum value for times earlier than $0.8 t_f$. This confinement of infall speeds to values smaller than the initial velocity dispersion suggests that line profiles from a collapsing starless core should not have the supersonic "infall wings" expected for cores collapsing onto an already formed point mass (Anglada et al 1987). These wings arise because the infall velocity onto a point mass varies with distance r as $r^{-1/2}$. Line profiles corresponding to such infall for similarity solutions of Larson (1969) and Penston (1969) and Shu (1977) are computed and discussed by Zhou (1992). We note that supersonic velocities are expected in the isothermal Larson-Penston similarity solution, but as in the pressure-free collapse discussed here, these supersonic velocities arise only at times very close to that of point mass formation. For nearly all of the infall phase of the collapse, both the pressure-free and isothermal collapse models have subsonic velocities.

## 3.3. Density structure

Figure 2 compares density structure of the same three centrally condensed collapsing systems as in Figure 1, at both the initial instant and after half a free-fall time. All three initial density profiles have "flat" profiles at the center and become steeper at larger coordinate values. The log-log slope approaches –2.5 for the sphere, -4.0 for the cylinder, and increases exponentially for the layer, reaching about –8 at z=0.05 pc. After half a free-fall time, the outer density structure of each system is nearly unchanged, but the peak densities have increased by a factor ~1.3, 1.6, and 2.0 for the layer, cylinder, and sphere respectively. This behavior indicates that it may be very difficult to assess the dynamical state of a centrally condensed starless core using observations sensitive only to density or column density, such as those based on dust absorption or emission. Although many cores have been well-fit with "Bonnor-Ebert" profiles corresponding to equilibrium states, Figure 2 shows that a core in the early stages of collapse would also have a similar profile. This near-degeneracy of static and collapsing density profiles contrasts with the dramatic change in power-law density profiles between the static and collapsing singular isothermal sphere (Shu 1977).

The similarity of density profiles of collapsing Bonnor-Ebert spheres can be even closer than in Figure 2 when the initial states are allowed to differ. In Figure 3a, the current state is chosen to have $n_0 = 3 \times 10^5$ cm$^{-3}$ at collapse ages $t/t_f = 0.0$, 0.4, 0.6, and 0.8, each with initial central density and velocity dispersion chosen to give a close match. The initial states are specified by matching the static and collapsing density profiles at zero and large radius, yielding

$$n_{0f} = n_{0s}\cos^6\beta(0) \qquad (22)$$

$$\sigma_f = \sigma_s\cos\beta(0) \qquad (23)$$

where we now use subscripts f and s to refer to the free-fall and static solutions, respectively. Here $\beta(0)$ is the value of $\beta$ at the origin, obtained from eqs. (5) and (9) via

$$\beta(0) + (1/2)\sin(2\beta(0)) = (\pi/2)(t/t_f) \qquad (24)$$

Figure 3a shows that for the first half of a central free-fall time, static and collapsing density profiles can be nearly identical except for a relatively small change in scale factor. Among the profiles for $t/t_f = 0$, 0.4, and 0.6, there is no deviation greater than about 30%. For all practical purposes these profiles would provide fits of virtually equal quality to a realistic set of observational data. This similarity of shape for early times can be easily understood since the inner part of the initial density profile is essentially uniform, and a collapsing uniform sphere gets denser and smaller but stays uniform. The outer part of the profile becomes slightly shallower in slope, but for early times this change is very small because the gas has hardly moved from its initial position.

In contrast, Figure 3b shows the velocity profiles for the same initial conditions and the same times $t/t_f = 0.4$, 0.6, and 0.8 as in Figure 3a. Here the collapse velocity profiles are clearly distinguishable from each other, and from the initial state where v=0 (not shown). These results are also understandable from the analysis of Section 3.1. The velocity profile maintains essentially the same shape, but the multiplying scale factor $t/t_f$ increases significantly, so the profile amplitude grows in proportion.

3.4. Uniform collapse

The collapse of an initially uniform sphere given by Hunter (1962) and of an initially uniform cylinder and layer given by Penston (1969) indicate that in each case the density remains spatially uniform as it increases with time; and that the infall speed increases linearly with increasing values of the spatial coordinate. At a given time, the maximum speed depends only on the size of the collapsing region. Thus systems which collapse from a uniform initial state will always have their largest infall velocities on the outside. These conclusions also apply to collapsing prolate and oblate ellipsoids, except that these do not maintain their axial ratios: as they collapse, uniform prolate objects become more prolate, and uniform oblate objects become more oblate (Lin, Mestel, & Shu 1965).

The uniform collapse results are useful for understanding velocity profiles in condensed systems whose density profiles are shallow on the inside and steep on the outside. For collapsing spheres, cylinders, and layers with such density profiles, the inward velocity increases outward from the center, just as in a uniform collapse. But at large radii, the shallow density profile causes the enclosed mass to increase very slowly with increasing radius. Then for spheres and cylinders the decrease of gravity with distance dominates the increase of gravity with enclosed mass, causing the velocity to decrease with distance. However for layers, the inward gravitational force does not depend on distance, so the velocity approaches its maximum value asymptotically at the largest distance, due solely to the increase in enclosed mass.

3.5. Mass accretion rates

The shallow inner density structure of the initial states considered here plays a crucial role in the evolution of the collapse, giving distinctly different "infall" and "accretion" phases. To illustrate this we use the collapse equations of section 2.1 to obtain the mass accretion rate of a collapsing Bonnor-Ebert sphere as a function of time since the onset of collapse.

The mass of the shell initially between $\xi$ and $\xi + d\xi$ can be written

$$dm = 4\pi\rho_0 \, a_0^3 \, \exp(-\psi)\xi^2 \, d\xi \qquad (25)$$

We call $t_f(\xi)$ the free-fall time for this shell, which is related to the central free-fall time $t_f$ in eq. (4) by

$$t_f(\xi) = t_f \, [\xi/(3\psi')]^{1/2} \qquad (26)$$

Taking the derivative of eq. (26) gives the time interval for this mass shell to cross the origin,

$$dt = t_f \, (3^{1/2}/2)(1-\mathscr{u}/3)\mathscr{v}^{1/2} \, d\xi \qquad (27)$$

where $\mathscr{u} \equiv (\xi/\psi') \exp(-\psi)$ and $\mathscr{v} \equiv \xi\psi'$ are dimensionless functions of $\xi$ tabulated by Chandrasekhar & Wares (1949). Note that $\mathscr{v}$ should not be confused with the velocity v. Then the central mass accretion rate is obtained from eqs. (25) and (27),

$$dm/dt = (\sigma^3/G)[\mathscr{u}\mathscr{v}^{3/2}]/[1-\mathscr{u}/3] \qquad (28)$$

Eq. (26) for the time since the start of collapse and eq. (28) for the central mass accretion rate each depend only on $\xi$, so the central mass accretion rate can be expressed in terms of time, as in Figure 4. For comparison Figure 4 also gives the central mass accretion rate for the pressure-free collapse of a singular isothermal sphere (SIS), which is obtained by similar analysis to be $dm/dt = (8/\pi)(\sigma^3/G)$. Note that this rate exceeds that of the isothermal collapse of a SIS (Shu 1977) by the factor 2.61.

Figure 4 shows that the central mass accretion rate is zero for the first central free fall time. During this "infall" phase the mass shells from the shallow-slope portion of the density profile travel together toward the origin, just as in the collapse of a uniform sphere. Then dm/dt increases sharply as these mass shells hit the origin close together in time, forming a point mass and starting the "accretion" phase. Finally, dm/dt declines gradually toward the constant rate of the SIS, as the mass shells from the outer, steep-slope portion of the density profile hit the origin at progressively later times. Similar results were found for non-equilibrium initial states by Henriksen, Andre′ & Bontemps (1997) in analyzing collapse of a density profile with an inner "flat-top", and by Whitworth & Ward-Thompson (2001) in their analysis of the collapse of a "Plummer-like" sphere.

Eq. (28) and Fig. 4 indicate that the mass accretion rate reaches a maximum of $7.07(\sigma^3/G)$, greater than for the constant rates of either the isothermal or pressure-free collapse of the SIS, and less than the maximum rate for either the isothermal collapse of the Bonnor-Ebert sphere computed numerically by Foster & Chevalier (1993), or the isothermal collapse of a uniform sphere, computed numerically by Larson (1969).

This temporal behavior of the collapsing Bonnor-Ebert sphere is sharply different from the collapse of any initial state whose density profile has a constant negative power law, including the SIS. For those cases, inner mass shells hit the origin from the first moment of collapse, so a point mass forms promptly and the collapse evolution is pure accretion with no preceding infall period.

These results indicate that "collapsing starless cores" can exist only when preceded by an initial state whose inner density profile has very shallow slope. The relative duration of the infall and accretion phases then depends on the relative extent of the shallow- and steep-slope portions of the density profile. The infall and accretion periods are in the ratios 1:1 and 1:2 for initial values $\xi_1=5.2$ and $\xi_1=8.2$, respectively. For the Bonnor-Ebert initial state with $\xi_1=10$ as shown in Figure 4, the infall and high-accretion-rate periods are similar, suggesting that collapsing starless cores and "class zero protostars" (André, Ward-Thompson & Barsony 1993) should have similar duration and thus similar numbers among nearby star-forming regions.

4. COMPARISON TO OBSERVATIONS

To apply the results in Section 3 we consider the observations of starless cores having centrally condensed density profiles and spatially extended infall asymmetry, as summarized in Section 1.

4.1. Initial geometry

Initially uniform collapse models can be ruled out as unlikely, since well-studied starless cores are centrally condensed, and uniform or nearly-uniform dense cores are not observed in star-forming regions.

Similar arguments tend to rule out initially centrally condensed layers and to favor initially centrally condensed cylinders and spheres. The vertically collapsing layer is a poor match to the typical observations. Even if the infinite layer is approximated by a finite layer, or by an oblate spheroid, to match observed central concentration the object must be viewed nearly edge-on, which is statistically unlikely. Even if it were viewed nearly edge-on, its inward motions would lie primarily in the plane of the sky, and thus would be difficult to detect. A radially collapsing cylinder is a better match, provided it is approximated by a finite cylinder or a prolate spheroid, and provided it is viewed nearly broadside. Then its elongated projected shape can match observed maps and its inward velocities can lie primarily along the line of sight. The collapsing sphere has no orientation constraints, and its inward velocities can also lie primarily along the line of sight, for sight lines passing near the center, but its projected shape does not match the typical 2:1 map elongation.

A radially collapsing disk can be considered a "slice" of a radially collapsing cylinder, and thus these two systems should have similar density profiles which are shallow on the inside and steep on the outside; and similar velocity profiles which are zero at the center, maximum near the center, and declining toward the outside. Even when the inward gravitational pull is opposed by thermal pressure and the frictional force of ion-neutral collisions, as in ambipolar diffusion models, these same properties of the velocity and density profiles are seen in numerical simulations (e.g. Ciolek & Mouschovias 1995, Ciolek & Basu 2000). Thus detectability of infall asymmetry puts the same requirements on orientation with respect to the line of sight for a radially collapsing disk (nearly edge-on) as for a radially collapsing cylinder (nearly broadside), but the projected elongation of the disk should then be perpendicular to its symmetry axis rather than along the symmetry axis as for the cylinder.

To further compare the collapse models of Section 2 with observations, we consider for simplicity the centrally condensed sphere. Since the radially collapsing infinite cylinder has a radial velocity profile similar to that of the collapsing sphere, it would also be useful in future work to consider in more detail the radial collapse of finite prolate objects, extending the work of Lin, Mestel & Shu (1965), Bastien (1983), and Rouleau & Bastien (1990).

4.2 Infall Age and Initial Condensation

The spherical infall model of Section 3 predicts the velocity and density profiles as a function of time. These in turn can be used with a radiative transfer model to predict line profiles which can be matched directly with observed lines. Such a detailed comparison is desirable, but is beyond the scope of this paper. Here we rely instead on matching the effective infall speed $v_{in}$, a less direct but simpler comparison.

The CS 2-1 observations of Lee, Myers & Tafalla (2001; hereafter LMT01) indicate that half the 19 cores analyzed have effective infall speed $v_{in}$ in the range 0.05 - 0.09 km s$^{-1}$ based on simple analytic models of the line profile. These same cores have infall asymmetry detected over projected radii 0.1-0.2 pc. To match these properties with the infall model one needs to specify the initial temperature $T_i$ and the initial central density $n_0$. We base these choices on properties of the cores L1517B and L1696A = Oph D, whose $N_2H^+$ lines are some of the narrowest known, among starless cores which have been searched for, but do not show, evidence of infall motions (LMT01). These cores appear most advanced in dissipating their turbulent motions, and thus may be in an evolutionary state close to the start of collapse. Their lines of $N_2H^+$ 1-0 have FWHM = 0.20 km s$^{-1}$ after correction for spectrometer broadening, corresponding to pure thermal broadening at temperature $T_i$ = 25 K. Their peak central densities are respectively $2 \times 10^5$ cm$^{-3}$, based on observations of millimeter wavelength dust emission (Tafalla et al 2004), and $3 \times 10^5$ cm$^{-3}$, based on observations of mid-infrared dust absorption (Bacmann et al 2000). We adopt $n_0 = 3 \times 10^5$ cm$^{-3}$.

With these choices of $T_i$ and $n_0$, the spherical infall model can constrain the collapse age once a suitable way is chosen to obtain the effective infall speed $v_{in}$ from the infall model for comparison with $v_{in}$ derived from the CS 2-1 line profile. Since the CS molecules are strongly depleted due to freeze-out in the inner part of the core (Tafalla et al 2002, 2004), we obtain $v_{in}$ by computing the mean infall speed, weighted by the gas density, for all densities between the threshold for depletion, $n_{dep}$, and the density at the outer radius. The result is that the observed range of infall speeds 0.05-0.09 km s$^{-1}$ corresponds to times $t/t_f$ = 0.3-0.5, and that the radius $r_{in}$, within which CS line infall asymmetry is observed corresponds to dimensionless radii $\xi > 10$, i.e. unstable against collapse in an equilibrium model.

We conclude that the spherical infall model is consistent with the typical range of core infall speeds and spatial extents of infall asymmetry observed in the CS 2-1 line, for initial central density $3 \times 10^5$ cm$^{-3}$ and initial FWHM line width 0.20 km s$^{-1}$, as observed in the narrow-line cores L1517B and Oph-D, provided that the initial radii correspond to unstable and highly condensed states, with $\xi > 10$, and that the collapse is relatively young, i.e. $t/t_f$ = 0.3-0.5.

4.3 Line Profiles

In this section we use the model of spherical infall in Section 3 to predict the time evolution of an optically thin line observed toward an initially condensed, collapsing sphere. These results indicate that as the collapse progresses, the thin line profile gets broader and eventually develops two peaks, whose separation approaches twice the maximum infall speed. This behavior provides a useful diagnostic of infall motions, and allows a simple way to discriminate among cores with infall asymmetry. In particular, this model indicates that among cores with evidence of inward motions, L1544, whose optically thin spectra show two peaks, is in a later stage of infall than other starless cores whose thin lines have only one peak.

We compute the optically thin emission profile by evaluating the optical depth $\tau$ for a frequently observed line of low optical depth, the hyperfine component of $N_2H^+$ 1-0 having the lowest statistical weight ($JF_1F$ = 110-011). We compute $\tau$ along the central line of sight z as a function of line-of-sight velocity V, according to

$$\tau = C \int_{-r_{max}}^{r_{max}} dz \, n \frac{x_{mol}}{T_{ex}} \exp\left[-\frac{1}{2}\left(\frac{V-v}{\sigma_{mol}}\right)^2\right] \qquad (29)$$

where C is a constant, $r_{max}$ is the maximum radius for which emission is computed, n is the gas density, $x_{mol}$ is the molecular abundance, $T_{ex}$ is the transition excitation temperature, $\sigma_{mol}$ is the velocity dispersion of the observed species, and v is the infall speed. The density, radius, and infall speed are given by eqs. (1) - (3), and the excitation temperature $T_{ex}$ is given in a two-level excitation model as

$$T_{ex} = T \frac{T_b + \frac{n}{n_{cr}} T_0}{T + \frac{n}{n_{cr}} T} \qquad (30)$$

(e.g., Rohlfs & Wilson 1999), where $n_{cr}$ is the critical density for excitation, assumed be $2 \times 10^5$ cm$^{-3}$ for $N_2H^+$ 1-0 (Ungerechts et al 1995), $T_b$ is the cosmic background temperature, 2.72 K, T is the isothermal kinetic temperature, assumed to be 10 K, and $T_0$ is the transition temperature, defined to be $T_0 = h\nu/k$, where h is Planck's constant, $\nu$ is the transition frequency, and k is Boltzmann's constant.

To evaluate the optical depth we assume as in Section 4.2 above that the initial equilibrium temperature $T_i$ is 25 K at times t < 0, corresponding to observed $N_2H^+$. line widths. Then at t=0 the gas is suddenly cooled to T=10 K, and the free-fall collapse ensues. The gas at T=10 K is assumed to contribute negligibly to the pressure which opposes collapse, but provides line broadening with $\sigma_{mol}=\sqrt{(kT/m_{mol})}$. The abundance $x_{mol}$ is assumed either to be constant, i.e. no freeze-out onto dust grains, or to vary with gas density n as $x_{mol}= \exp(-n/n_{dep})$ where $n_{dep}$ has a value $3 \times 10^4$ cm$^{-3}$ typical for CS in isolated low-mass cores studied by Tafalla et al (2002, 2004).

The computed line profiles are shown in Figure 5, for times t<0, and for $t/t_f$=0, 0.3, 0.4, and 0.5. At the early times t <0 and t=0, the line profiles are single-peaked Gaussians, reflecting the assumed initial conditions. At the later times, the lines broaden and develop two peaks as the characteristic inward speed becomes comparable to the molecular velocity dispersion. The two-peak structure first becomes noticeable at about $t/t_f = 0.4$. For the times t < 0 and t = 0 there is no difference between the profiles with and without depletion. For $t/t_f = 0.3$, 0.4, and 0.5 the depleted profiles have somewhat deeper dips than the undepleted profiles.

The broadening and splitting of optically thin lines from a collapsing starless core can be understood in terms of the ratio of the maximum infall speed $v_{max}$ to the molecular velocity dispersion $\sigma_{mol}$. At the sequence of times in Figure 5, $t/t_f = 0.3$, 0.4, and 0.5, the ratio $(v_{max}/\sigma_{mol})$ has respective values 1.5, 2.0, and 2.5. Thus the line profile splits into two peaks, when the collapse age satisfies

$$\frac{t}{t_f} \approx 0.4 \qquad (31)$$

or, using eq. (20), when the ratio of speeds becomes

$$\frac{v_{max}}{\sigma_{mol}} \approx 0.4\sqrt{\frac{T_i m_{mol}}{T m}} \qquad (32)$$

Eq. (32) implies that species with higher molecular mass, such as CS, CCS, and HC$_3$N, are better able to show two peaks than species with lower molecular mass, because the species with higher $m_{mol}$ have lower thermal molecular velocity dispersion $\sigma_{mol}$.

The effect of freeze-out with $n_{dep} = 3 \times 10^4$ cm$^{-3}$ is to deepen the dip between the profile peaks. This can be understood since freeze-out preferentially depletes molecules from the densest gas, and in the infall phase of the collapse, the densest gas has the lowest velocities. As $t/t_f$ increases, $n/n_{dep}$ increases, causing more molecules to deplete, deepening the dip. We caution that freeze-out can also significantly reduce the amplitude of the profile, but we do not display this result here.

On the basis of these considerations, we have examined the 31 starless cores in the N$_2$H$^+$ 1-0 survey of Crapsi et al (2004), for profile features expected in the infall model presented here: thin 2-peak profiles, lines with unusually large widths, and lines whose width increases toward the center of the map. In this group, only L1544 clearly shows all three of these properties. Furthermore, two-peak thin profiles are seen toward L1544 in other tracers: HC$^{18}$O$^+$ 1-0 and H$^{13}$CO$^+$ 1-0 (Caselli et al 1999) and CCS $J_N = 3_2$-$2_1$ (Ohashi et al 1999). Among the other notable cores, L429 has an unusually broad profile with a suggestion of two peaks, but its line width does not increase inward. L694-2 has inward-increasing line widths, but they are not unusually large, and the thin N$_2$H$^+$ line profiles show at most a slight red shoulder rather than two distinct peaks. L328 shows an unusually broad line, but with only one peak. We note that line broadening can arise from turbulence, rotation, and outflow in addition to infall, while the presence of two peaks with the velocity separation discussed above may be a feature more specific to infall. In conclusion, only L1544 shows this evidence of "advanced infall."

## 5. DISCUSSION

### 5.1 Core condensation and early collapse

Extended infall asymmetry has been interpreted as reflecting the gradual motions of core condensation, and thus as part of the initial conditions for star-forming collapse (Tafalla et al 1998, Lee, Myers & Tafalla 2001). Similarly, it has been described as reflecting the late stages of ambipolar diffusion, in a system where the mass-to-flux ratio is nearly critical, for L1544 (Ciolek & Basu 2000). In this paper we have shown that extended inward motions may also be consistent with the early stages of pressure-free collapse from an initially condensed isothermal equilibrium state, provided there is sufficient initial condensation. The motions agree in some detail with collapse of an unstable Bonnor-Ebert sphere, and also agree qualititatively with collapse of an isothermal cylinder.

Another, closely related interpretation is that the extended inward motions, regardless of their physical origin, do not directly reflect gravitational collapse, but nonetheless immediately precede the collapse as part of its "initial conditions." This situation has been explored in detail for the spherical collapse problem, in the context of isothermal and polytropic equations of state, with and without initial overdensity (Fatuzzo, Adams, & Myers 2004). The main conclusion is that if extended inward motions with velocities of about half the sound speed immediately precede collapse, they speed up the collapse by about a factor of two.

These two processes - late stages of core condensation and early stages of collapse - refer to nearly the same period of time in the evolution of a star-forming core, but early collapse is a simpler and less ambiguous explanation.

The physical basis of core condensation is still poorly understood. It has often been described as a slow or quasi-static process (Shu, Adams & Lizano 1987). If this description is correct, the basic physics is still unclear, since gravity can be opposed by ion-neutral friction (ambipolar diffusion; Mestel & Spitzer 1954), or by dissipating turbulent pressure (Nakano 1998, Myers & Lazarian 1998), by turbulent ambipolar diffusion (Zweibel 2002, Fatuzzo & Adams 2002), or by other processes. It has also been suggested that the condensation which precedes star-forming collapse may operate more quickly, via supersonic turbulent flows (Elmegreen 1993, Padoan, Nordlund & Jones 1997, MacLow & Klessen 2004).

In contrast, gravitational infall is a well-defined model, especially in the simple form considered here. Furthermore the initial states of the collapse models of section 4 which best match the observed core data are slightly less dense versions of these observed cores, but with no inward motions. Many of the ~50 starless cores surveyed for extended infall asymmetry but which show no clear sign of infall asymmetry (Lee, Myers & Tafalla 2001), match this description.

The polarization and Zeeman measurements discussed in Section 1 indicate that magnetic fields are present in cores, so in a full treatment they must be included. We speculate that if magnetic forces are neither negligible nor dominant, their main effect will be to slow down the collapse by a factor comparable to or less than that due to thermal pressure, discussed in the next subsection. Although a magnetically diluted collapse model appears consistent with two-peaked optically thin lines in L1544 (Caselli et al 2002), the results in section 4.3 indicate that a nonmagnetic collapse model can also reproduce such lines.

5.2. Pressure-free and isothermal collapse

The pressure-free collapse calculations presented here assume implicitly that the thermal pressure is negligible during the collapse. This is unrealistic since observations indicate that most cores in regions of low-mass star formation, including those with evidence of inward motions, have gas kinetic temperatures close to 10 K (Jijina, Myers, & Adams 1999). A more realistic calculation would include the effects of thermal pressure, as was done in the numerical calculations of collapse of a Bonnor-Ebert sphere by Foster & Chevalier (1993) by Ogino, Tomisaka & Nakamura (1999), and by Aikawa et al (2004).

The analytic pressure-free collapses reported here have the same qualitative features as the numerical isothermal collapses cited above, over the first free-fall time--density profiles with shallow inner slope and steep outer slope, and velocity profiles with a local maximum. The differences between the isothermal and pressure-free solutions are minimal at the start of collapse, and become more pronounced by the time of point mass formation. The isothermal collapse is generally slower than the corresponding pressure-free collapse, and the duration of the isothermal collapse depends on how the collapse is initiated. The typical ratio of isothermal and pressure-free durations is a factor of a few. The central time scale for isothermal collapse is longer than that of pressure-free collapse by a factor 1.58, according to the Larson-Penston similarity solution (Larson 1969). The mass accretion rate reaches its peak value about 1.2 times later for the isothermal collapse than for the pressure-free collapse, for the same starting conditions (Motoyama & Yoshida 2003). If the initial overdensity factor for isothermal collapse is 1.1, the ratio of collapse time to central free-fall time is greater by a factor 3.4 than when the initial overdensity factor is 4.0, corresponding to nearly pressure-free collapse (Yukawa et al 2004).

Thus the analytic results for early collapse given here provide a qualitative guide to the more realistic isothermal collapse, which must be computed numerically. The analytic predictions of collapse times appear to underestimate the corresponding numerical times by a factor of order a few, but a dedicated numerical study is required to obtain a more detailed estimate of the accuracy of the analytic predictions.

## 5.3. Initiation of collapse

The foregoing description of cores collapsing from centrally condensed initial states begs the questions of how such states are reached, and how the collapse is initiated. Many suggestions have been made for each of these two points, including quasistatic approaches such as ambipolar diffusion (Mestel & Spitzer 1956, Mouschovias 1976, Ciolek & Basu 2000) and turbulent dissipation (Nakano 1998, Myers & Lazarian 1998), and faster processes such as shocks driven by stellar winds (Boss 1995) or supernovae (Vanhalla & Cameron 1998), or converging turbulent flows (Ballesteros-Paredes, Hartmann, & Vazquez-Semadeni 1999). Here we note that any of these processes which can provide a sufficiently centrally condensed structure with relatively low velocity dispersion may be viable. A detailed comparison of these processes would be a useful subject for future studies.

## 5.4. Centrally condensed collapse

The initial states for collapse discussed in this paper are all idealized isothermal equilibria. These were chosen because they have well-defined analytic density distributions which have zero initial velocity. Collapse from these initial states is a mathematically convenient way to achieve the observed density and velocity profiles, but not the only way, for several reasons.

First, the initial states need not be strict equilibrium states in hydrostatic force balance. The "Plummer-like" density profile proposed by Whitworth & Ward-Thompson (2001) has similar shape to the Bonnor-Ebert profile, but it is not a hydrostatic equilibrium structure. Still, during early collapse it gives velocity and density profiles similar to those from the condensed Bonnor-Ebert collapse discussed here. Second, examination of the collapse solution of Hunter (1962) indicates that nonzero initial inward velocities reduce the time scale but do not significantly change the structure of the density and velocity profiles during collapse, as long as the initial velocities are small compared to the initial sound speed. A similar conclusion is reached by Adams, Fatuzzo & Myers (2004) for the similarity solutions to the collapsing singular isothermal sphere. Third, the similarity of the density and velocity profiles between the collapsing cylinder and the collapsing condensed sphere, shown in Fig. 2, suggests that collapse from centrally concentrated elongated objects with a range of aspect ratios and shapes may lead to similar velocity profiles and similar density profiles.

Thus the essential property of the initial state for collapse, needed to match observed density and velocity profiles, appears to be strong central condensation, rather than strict hydrostatic equilibrium, or zero initial velocity, or strict spherical shape.

5.5. Summary

In this paper models are presented of free-fall gravitational collapse from nonsingular isothermal equilibrium states of spheres, cylinders and layers. These initial states are more realistic models of starless dense cores than either uniform or singular initial states, and the resulting early-stage properties of the collapsing systems differ from each of these well-known solutions. The main conclusions are:

1. For times earlier than the central free-fall time, velocity profiles have a maximum value, approximately the initial sound speed times the ratio of the collapse age to the central free-fall time. This maximum velocity occurs near the initial scale height for cylinders and for centrally condensed spheres, but at the outermost radius for layers and for lightly condensed spheres.

2. Density profiles of these same collapsing systems have shallow inner slopes and steep outer slopes, and are difficult to distinguish from density profiles of isothermal equilibrium. For the sphere whose initial state is an equilibrium Bonnor-Ebert sphere and whose collapse age is half a central free-fall time, the collapse density profile deviates by less than ~30% from a best-fit equilibrium density profile.

3. The maximum velocity for early collapse from isothermal equilibrium is subsonic, so spectral line profiles from such collapsing cores should not show the "collapse wings" expected for later stages of star formation, after a point mass has formed.

4. The collapse history depends crucially on the initial density profile. For a shallow inner slope and steep outer slope, as expected in isothermal equilibrium, the collapse has distinct "infall" and "accretion" phases of comparable duration, suggesting that there should be comparable numbers of "collapsing starless cores" and "class zero protostars." If the density profile has a constant negative power-law exponent, as in the case of the singular isothermal sphere, then no infall phase is expected.

5. Collapse models are more consistent with observations of "infall asymmetry" around dense cores if their initial states are centrally condensed cylinders or spheres, and less consistent if their initial states are layers or lightly condensed spheres.

6. Gravitational collapse from strongly condensed and moderately elongated initial states appears quantitatively consistent with the extended infall asymmetry observed around starless cores. Spherical models are consistent with observed inward speeds 0.05-0.09 km s$^{-1}$ over 0.1-0.2 pc, provided the collapse has a typical age 0.3-0.5 $t_f$, and began from a centrally condensed state with dimensionless radius $\xi_{max} > 10$.

7. Optically thin line profiles from collapsing starless cores become broader with time and develop two peaks, as the systematic infall speed becomes comparable to the thermal velocity dispersion of the observed species. This feature indicates that L1544 has achieved higher infall speeds than other starless cores with evidence of inward motion.

8. Many observed features of starless cores with infall asymmetry are matched by well-defined models of gravitational collapse. Alternatively, models of the late stages of core formation, including ambipolar diffusion, turbulent dissipation, and dynamical turbulent flows, are not ruled out, but they are more complex and have more uncertainty.

9. The pressure-free collapse models presented here are a useful guide to more realistic isothermal collapse models, provided the collapse is still in its early stages, during roughly the first half of a central free-fall time.

10. The initial states of the collapsing core models which match observed properties are structures in isothermal equilibrium, but the necessary feature of the initial state is not its equilibrium but rather its nonsingular central condensation. The velocity and density structure of the collapse solutions which best match the observations can also arise from prolate or spherical structures which are centrally condensed, even if they do not satisfy force balance, and even if their initial velocities are not zero.

We thank an anonymous referee for useful comments and corrections. We thank Fred Adams, Tyler Bourke, Andi Burkert, Paola Caselli, Antonio Crapsi, Charlie Lada, Chris McKee, Chang Won Lee, Ralph Pudritz, Frank Shu, Steve Stahler, Mario Tafalla, Malcolm Walmsley, and Ellen Zweibel for helpful discussions. This work was supported by the NASA Origins of Solar Systems Program Grant NAG5-13050. We thank the Miller Institute for Basic Research for their support and the Astronomy Department of the University of California, Berkeley for their hospitality during the writing of this article.

REFERENCES

Adams, F. C., Fatuzzo, M., & Myers, P. C. 2004, ApJ, in press

Aikawa, Y., Herbst, E., Roberts, H., & Caselli, P. 2004, submitted to ApJ

Alves, J. F. , Lada, C. J., Lada, E. A. 2001, Nature, 409, 159

Andre′, P., Ward-Thompson, D., & Barsony, M. 1993, ApJ 406, 122

Andre′, P., Ward-Thompson, D., & Barsony, M. 2000, in Protostars and Planets IV, eds V. Mannings, A. Boss, & S. Russell (Tucson: University of Arizona Press), p. 59

Andre′, P., Ward-Thompson, D., & Motte, F. 1996, A&A, 314, 625.

Anglada, G., Rodriguez, L. F., Canto, J., Estalella, R., & Lopez, R. 1987, A&A 186, 280

Bacmann, A., Andre′, P., Puget, J.-L., Abergel, A., Bontemps, S., & Ward-Thompson, D. 2000, A&A 361, 555

Ballesteros-Paredes, J., Hartmann, L., & Vazquez-Semadeni, E. 1999, ApJ 527, 285

Bastien, P. 1983, A&A 119, 109

Beichman, C. A., Myers, P. C., Emerson, J. P., Harris, S., Mathieu, R., Benson, P. J., & Jennings, R. E. 1986, ApJ, 307, 337

Benson, P. J., & Myers, P. C. 1989, ApJS, 71, 89

Bergin, E. A., & Langer, W. D. 1997, ApJ 486, 316

Bonnor, W. 1956, MNRAS, 116, 351

Boss, A. P. 1995, ApJ 439, 224

Bourke, T. L., Crapsi, A., Lee, C. W., Myers, P. C., Tafalla, M., & Wilner, D. J. 2004, in preparation

Caselli, P., Walmsley, C. M., Tafalla, M., Dore, L., & Myers, P. C. 1999, ApJL 523, 165

Caselli, P., Benson, P. J., Myers, P. C., & Tafalla, M. 2002a, ApJ, 572, 238

Caselli, P., Walmsley, C. M., Zucconi, A., Tafalla, M., Dore, L., & Myers, P. C. 2002b, ApJ 565, 331

Cernicharo, J. 1991, in The Physics of Star Formation and Early Stellar Evolution, eds. C. J. Lada & N. D. Kylafis (Dordrecht: Kluwer), p. 287

Chandrasekhar, S. 1939, An Introduction to the Study of Stellar Structure (Chicago: University of Chicago Press)

Chandrasekhar, S., & Wares, G. 1949, ApJ 109, 551

Ciolek, G. E., & Basu, S. 2000, ApJ, 529, 925

Ciolek, G. E., & Mouschovias, T. Ch. 1995, ApJ 454, 194

Crapsi, A., Caselli, P., Walmsley, C. M., Myers, P. C., Tafalla, M., Lee, C. W., & Bourke, T. L. 2004, in preparation

Crutcher, R. M., Nutter, D. J., Ward-Thompson, D., & Kirk, J. M. 2004, ApJ 600, 279

Crutcher, R. M., & Troland, T. H. 2000, ApJL 537, 139

De Vries, C., & Myers, P. C. 2004, in preparation

Ebert, R. 1955, Z. f Astr. 37, 217

Elmegreen, B. G. 1993, ApJL, 419, 29

Evans, N. J. II, Rawlings, J. M., Shirley, Y. L., & Mundy, L. G. 2001, ApJ, 557, 193

Fatuzzo, M., & Adams, F. C. 2002, ApJ 570, 210

Foster, P., & Chevalier, R. 1993, ApJ 416, 303

Fuller, G. A., & Myers, P. C. 1993, ApJ 418, 273

Goldsmith, P. F. 2001, ApJ, 557, 736

Goodman, A. A., Barranco, J. A., Wilner, D. J., Heyer, M. H. 1998, ApJ, 504, 223

Gregersen, E. M., & Evans, N. J. II 2000, ApJ 538, 260

Harvey, D. W., Wilner, D. J., Lada, C. J., Myers, P. C., & Alves, J. F. 2003, ApJ 598, 111

Henriksen, R., Andre′, P., & Bontemps, S. 1997, A&A 323, 549

Hunter, C. 1962, ApJ 136, 594

Hunter, C. 1977, ApJ 218, 834

Johnstone, D., Fich, M., Mitchell, G. F., & Moriarty-Schieven, G. 2001, ApJ 559, 307

Jijina, J., Myers, P. C., & Adams, F. C. 1999, ApJS 125, 161

Jones, C. E., Basu, S., & Dubinski, J. 2001, ApJ 551, 387

Larson, R. B. 1969, MNRAS, 145, 271

Lee, C. W., & Myers, P. C. 1999, ApJS 123, 233

Lee, C. W., Myers, P. C., & Tafalla, M. 1999, ApJ, 526, 788

___________ 2001, ApJS, 136, 703

Lee, C. W., Myers, P. C., & Plume, R. 2004, ApJS, in press

Li, Z., & Shu, F. H. 1996, ApJ 473, 873

Lin, C. C., Mestel, L., & Shu, F. H. 1965, ApJ 142, 1431

Mac Low, M., & Klessen R. 2004, RvMP 76, 125

Mestel, L., & Spitzer, L. 1956, MNRAS 116, 503

Motoyama, K., & Yoshida, T. 2003, MNRAS, 344, 461

Mouschovias, T. Ch. 1976, ApJ 207, 141

Myers, P. C. 1983, ApJ 270, 105

Myers, P. C., Mardones, D., Tafalla, M., Williams, J. P., & Wilner, D. J. 1966, ApJL 465, 133

Myers, P. C., & Benson, P. J. 1983, ApJ 266, 309

Myers, P. C., & Fuller, G. A. 1992, ApJ 396, 631

Myers, P. C., Fuller, G. A., Goodman, A. A., & Benson, P. J. 1991, ApJ 382, 555

Myers, P. C., & Lazarian, A. 1998, ApJL 507, 157

Nakano, T. 1998, ApJ 494, 587

Ogino, S., Tomisaka, K., & Nakamura, F. 1999, PASJ 51, 673

Ohashi, N., Lee, S., Wilner, D. J., & Hayashi, M. 1999, ApJL 518, 410

Onishi, T., Mizuno, A., & Fukui, Y. 1999, PASJ 51, 2570

Ostriker, J. 1964, ApJ 140, 1056

Padoan, P., Nordlund, A., & Jones, B. 1997, MNRAS 288, 145

Penston, M. 1969, MNRAS 144, 425

Rouleau, F., & Bastien, P. 1990, ApJ 355, 172

Ryden, B. 1996, ApJ 471, 822

Shu, F. H. 1977, ApJ 214, 488

Shu, F. H., Adams, F. C., & Lizano, S. 1987, ARA&A 25, 23

Spitzer, L. 1942, ApJ 95, 329

Spitzer, L. 1965, in Stars and Stellar Systems, Vol. VII, ed. G. P. Kuiper (Chicago: U. Chicago Press), p. 1

Stodolkiewicz, J. S. 1963, AcA 13, 30

Tafalla, M., Mardones, D., Myers, P. C., Caselli, P., Bachiller, R., & Benson, P. J. 1998, ApJ 504, 900

Tafalla, M., Myers, P. C., Caselli, P., Walmsley, C. M., & Comito, C. 2002, ApJ 569, 815

Tafalla, M., Myers, P. C., Caselli, P., & Walmsley, C. M. 2004, A&A 415, 191

Vanhalla, H., & Cameron, A. G. 1998, ApJ 508, 291

Ward-Thompson, D., Motte, F., & André, P. 1999, MNRAS 305, 143

Ward-Thompson, D., Kirk, J. M., Crutcher, R. M., Greaves, J. S., Holland, W. S., & André, P. 2000, ApJL, 545, 121

Ward-Thompson, D., Scott, P. F., Hills, R. E., & Andre′, P. 1994, MNRAS 267, 141.

Ward-Thompson, D., Andre′, P., & Kirk, J. M. 2002, MNRAS 329, 257

Whitworth, A., & Summers, D. 1985 MNRAS, 214, 1

Whitworth, A., & Ward-Thompson, D. 2001, ApJ 547, 317

Zhou, S. 1992, ApJ 394, 204

Zucconi, A., Walmsley, C. M., & Galli, D. 2001, A&A 376, 650

Zweibel, E. 2002, ApJ 567, 962

FIGURE LEGENDS

Figure 1 - Velocity profiles for a sphere, cylinder, and layer collapsing in free fall from initial isothermal equilibrium, with central density $2 \times 10^5$ cm$^{-3}$ and temperature 15 K. Each profile is evaluated at a collapse age equal to half a central free-fall time. The initial state of the sphere is a critically stable Bonnor-Ebert sphere with dimensionless outer radius $\xi_1 = 10$. The coordinate x represents the radius of the sphere, the radius of the cylinder, and the height of the layer, as appropriate. Each velocity profile is zero at the origin and reaches a maximum value near the origin for the sphere and cylinder, and at the outermost radius for the layer.

Figure 2 - Density profiles for the same sphere, cylinder, and layer as in Figure 1, at the initial instant and after half a central free-fall time. Each system maintains a similar shape with shallow profile at small radius and a steep decline at large radius, increasing its central density by a factor 1.3-2.

Figure 3 - (a), Density profile for a static Bonnor-Ebert sphere with temperature 10 K and central density $3 \times 10^5$ cm$^{-3}$, and for collapsing Bonnor-Ebert spheres at 0.4, 0.6, and 0.8 of a central free-fall time, each with initial central density and temperature chosen to give the same present-day density as the static sphere at small and large radius. The profiles are nearly identical in shape. (b), Velocity profiles for the collapsing spheres in (a), each showing maximum velocity

approximately equal to the initial velocity dispersion times the ratio of the collapse age and the central free-fall time, and each easily distinguished from the other collapse cases and from the static case of zero velocity.

Figure 4 - Central mass accretion rate dm/dt of the collapsing Bonnor-Ebert sphere, indicating an "infall" phase with $dm/dt = 0$ for the first central free-fall time, and an "accretion" phase with large dm/dt for later times after the point mass has formed. This behavior contrasts with the constant dm/dt of the singular isothermal sphere, which has no infall phase and which starts its accretion phase from the first moment of collapse.

Figure 5 - Normalized profiles of optical depth, equivalent to optically thin emission line profiles, predicted for the $JF_1F = 110$-$011$ line of $N_2H^+$, according to the spherical collapse model. At times $t < 0$, the initial equilibrium state has central density $n_0 = 3 \times 10^5$ cm$^{-3}$, outer dimensionless radius $\xi_{max} = 10$, and temperature $T_i = 25$ K corresponding to observations of cores with FWHM $N_2H^+$ line width 0.20 km s$^{-1}$. At $t = 0$ the core is "suddenly cooled" to 10 K and the resulting line profile is computed for times $t/t_f = 0.3, 0.4,$ and 0.5. The profiles are shown as solid curves, assuming that the $N_2H^+$ molecules have constant abundance, and as dotted curves, assuming that the $N_2H^+$ abundance varies as $\exp(-n/n_{dep})$, where $n_{dep} = 3 \times 10^4$ cm$^{-3}$, due to freeze-out onto grains. For $t/t_f > 0.3$ the line profile broadens and develops two peaks, as is seen in L1544.

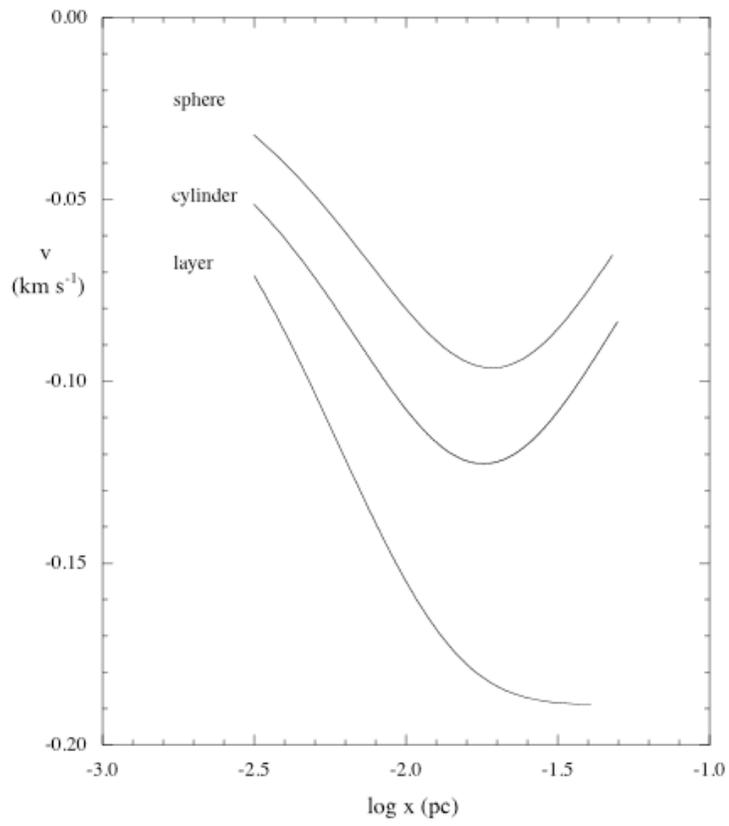

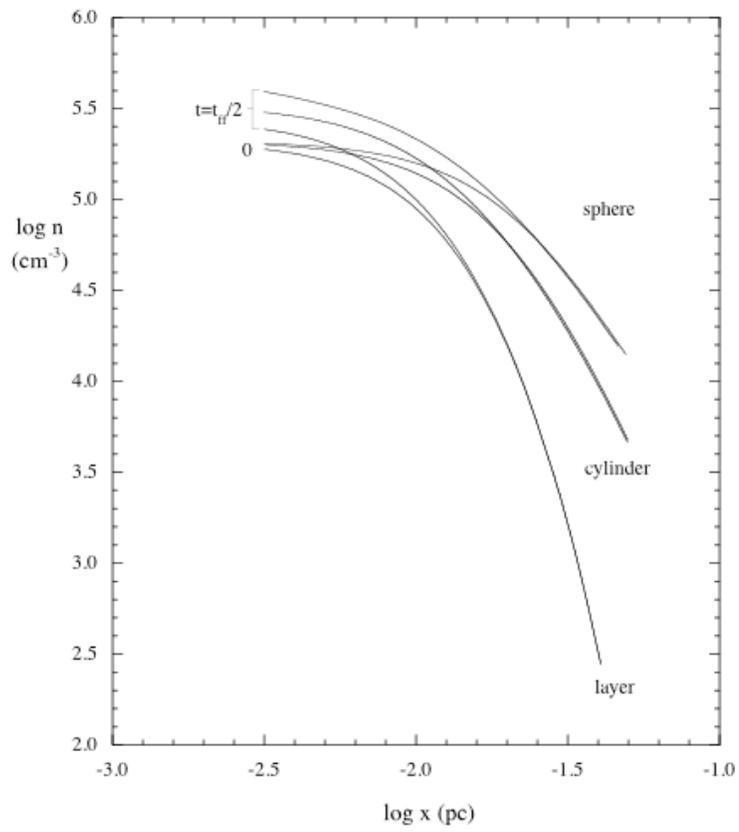

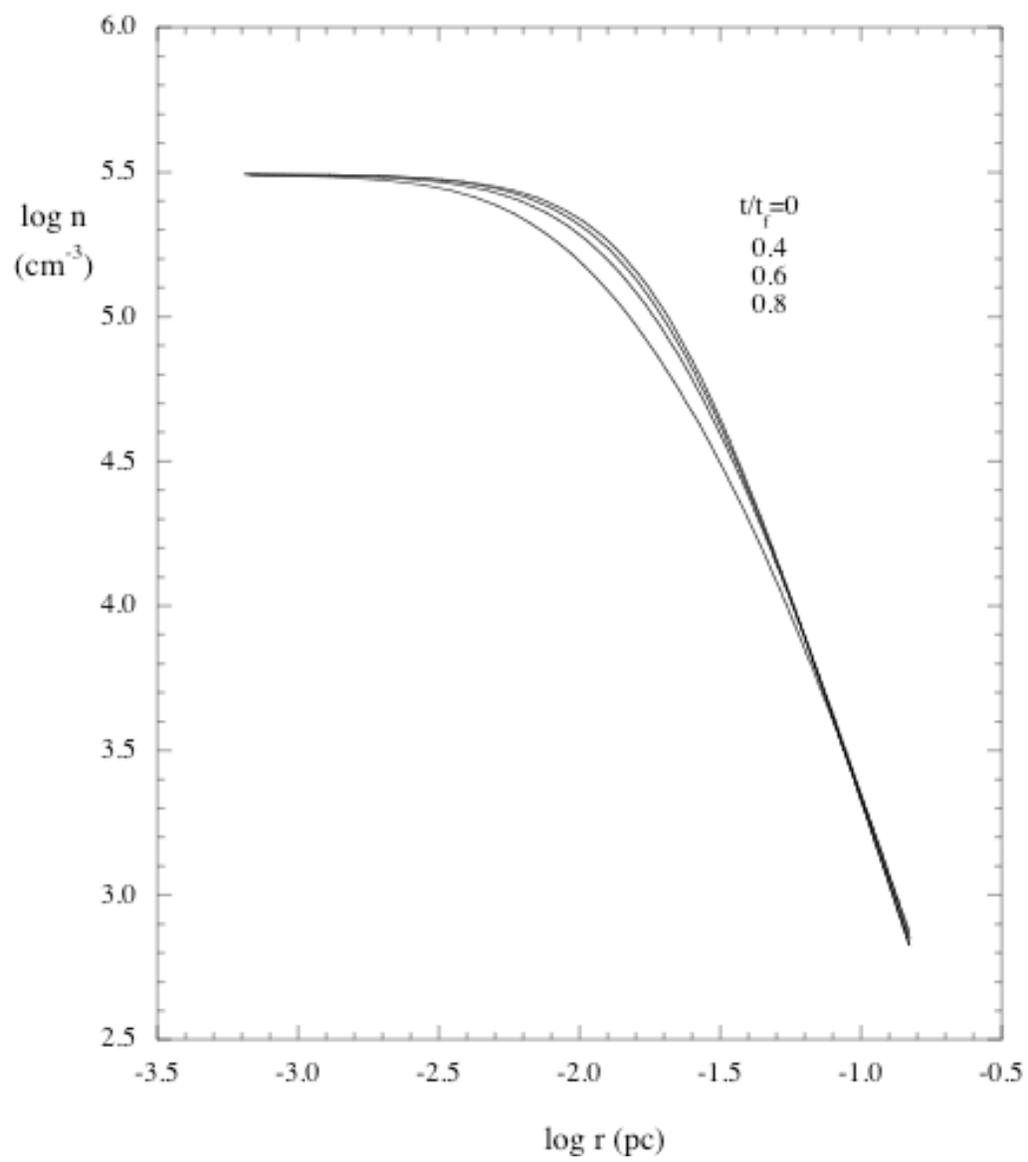

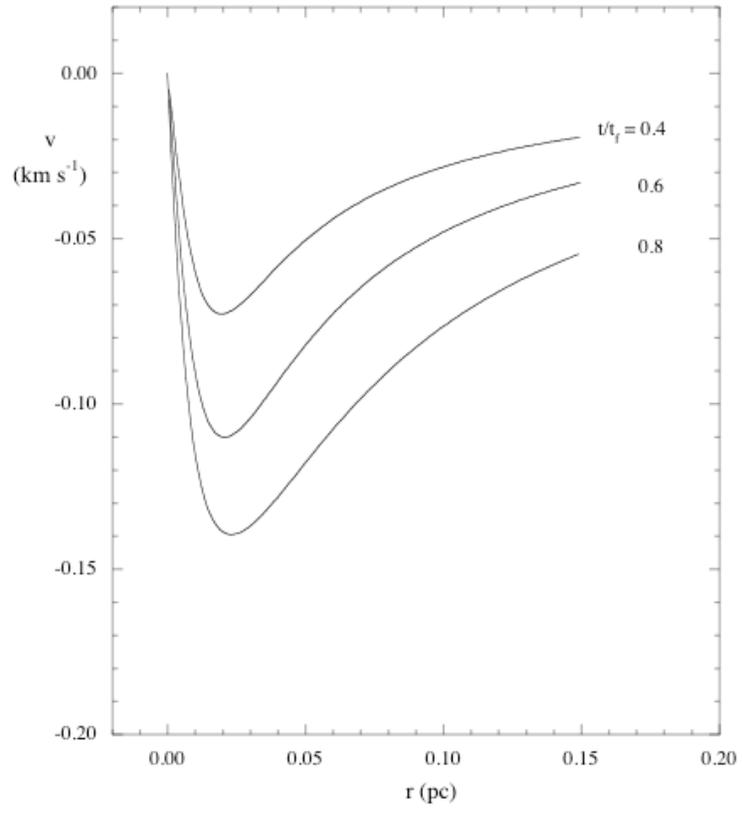

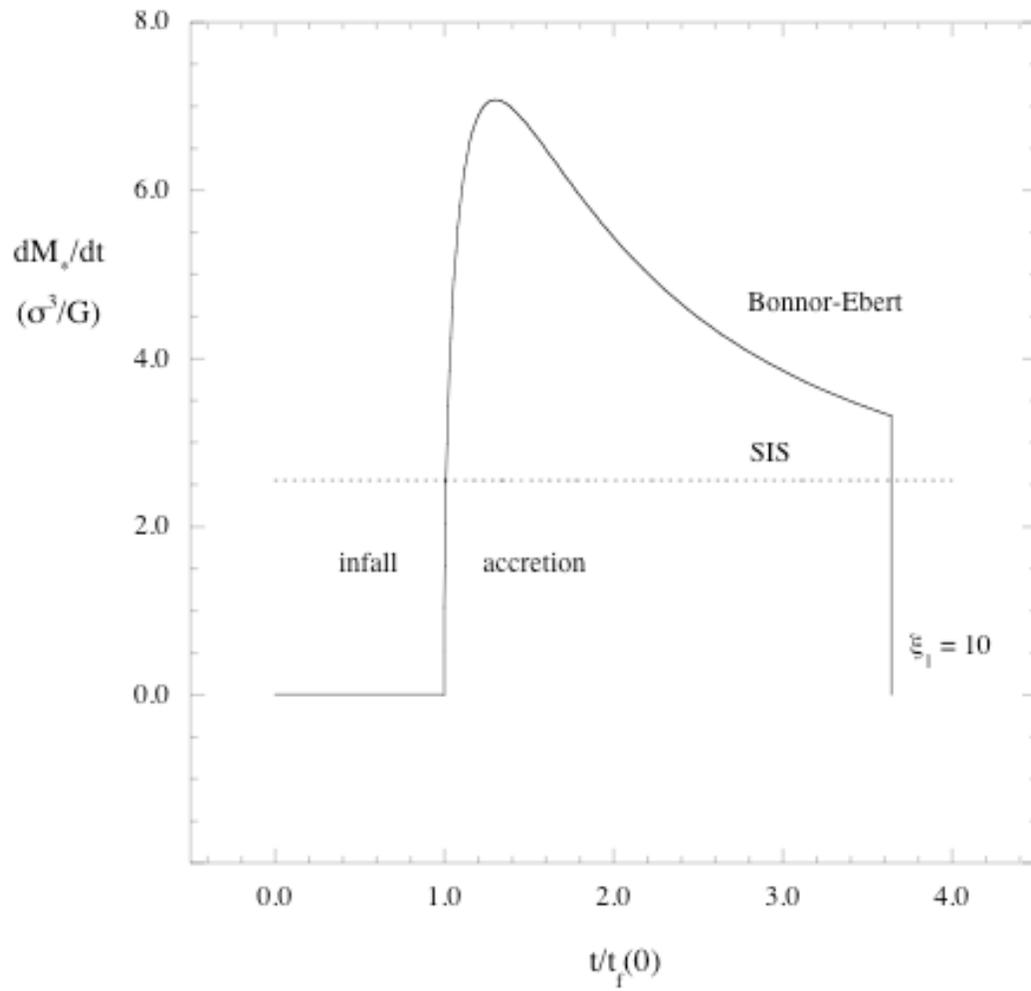

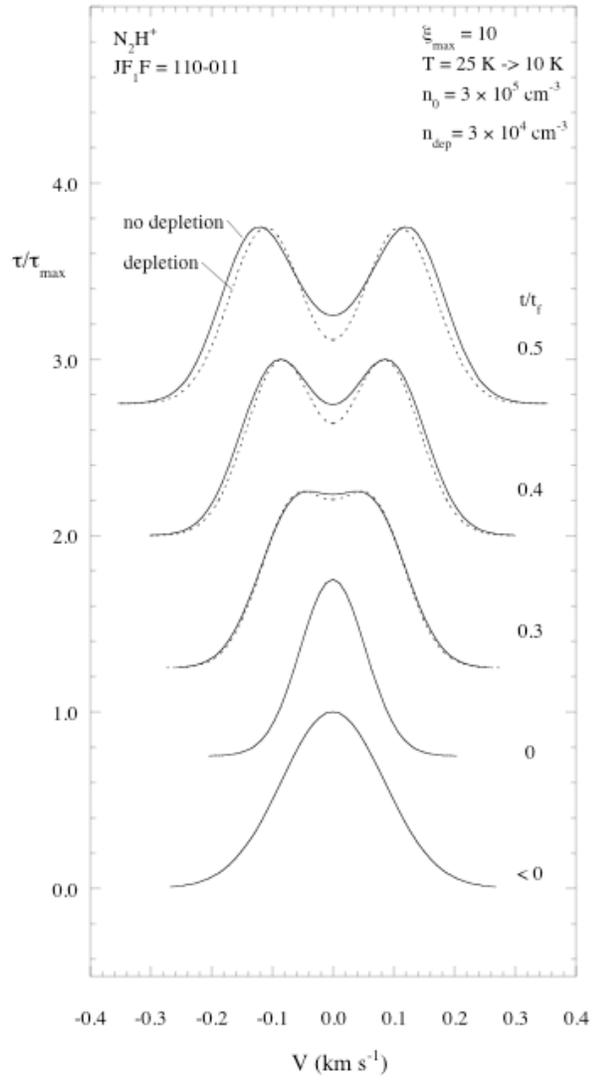